\documentclass[a4paper]{jpconf}
\usepackage{epsfig,rotating}
\usepackage{amsmath}
\usepackage{amsfonts}
\usepackage{amssymb}
\usepackage{graphicx}

\begin{document}
\title{Medium effects in direct reactions}

\author{M. Karakoc$^{(1,2)}$, and C. Bertulani$^{(1)}$}

\address{$^{(1)}$Department of Physics and Astronomy, Texas A$\&$M University-Commerce, Commerce, Texas 75429-3011, USA\footnote{mesut.karakoc@tamuc.edu}}

\address{$^{(2)}$Department of Physics, Akdeniz University, TR-07058 Antalya, Turkey\footnote{carlos.bertulani@tamuc.edu}}

\begin{abstract}
We discuss medium corrections of the nucleon-nucleon (NN) cross sections and their influence on 
direct reactions at intermediate energies $\gtrsim 50$ MeV/nucleon. The results obtained with free NN cross sections are compared with those obtained 
with a geometrical treatment of Pauli-blocking and with NN cross sections obtained with Dirac-Bruecker methods.
We show that medium corrections may lead to sizable 
modifications for collisions at intermediate energies and that they are more pronounced in reactions involving weakly bound nuclei.
\end{abstract}

\section{Introduction}
Obtaining a valid optical potential for direct reactions is very crucial for reactions such as nucleon knockout 
at intermediate and high energies~\cite{HRB91} ($\gtrsim 50$ MeV/nucleon). A microscopic method to deduce optical potentials is based on the construction of 
the potentials using an effective nucleon-nucleon (NN) interaction, or cross section (e.g. those of Ref. \cite{Ra79}). 
This technique  is often used to construct  the real part of an optical 
potential and with its imaginary part assumed  rescaled in strength to 
better reproduce experimental data on elastic scattering, or total reaction cross sections. The real and imaginary parts of the potential can also be constructed independently as 
in Refs. \cite{trache01,trache02}, where the procedure starts from a NN effective interaction with 
independent real and imaginary parts. It has also been shown that one can use nucleon-nucleon cross sections 
as the microscopic input \cite{HRB91}, instead of nucleon-nucleon interactions. In this case, an effective 
treatment of Pauli-blocking on nucleon-nucleon scattering is needed, as it manifests through medium 
density dependence. In fact, it is well known that a complete numerical modeling of 
heavy-ion central collision dynamics requires to  account for medium effects on 
the nucleon-nucleon cross sections \cite{Ber01}. The main goal in central collisions is to explore  the 
nuclear equation of state (EOS) by studying global collective variables describing the collision process. 

Medium modifications of NN scattering have smaller effects in direct reactions since generally low nuclear 
densities are probed. Although, no comparison with experimental data was supplied, 
a first work on this effect in knockout reactions was presented in Ref. \cite{BC10}. In this contribution, we report 
recent progress on studies of medium modifications in knockout reactions. We will report on medium effects in the NN cross section 
for the description of knockout reactions by means of (a) a geometrical treatment 
of Pauli-blocking and  a (b) Dirac-Brueckner treatment. A comparison of our calculations  to a large number of published 
experimental data is shown, and full results will be published else where \cite{KBBT12}. The aim of this project 
is to obtain more accurate spectroscopic factors that will lead to better understanding nuclear structure  
and to check and improve the credibility of the use of knockout reactions as an indirect methods for nuclear 
astrophysics.

\section{Medium effects}
\subsection{Nucleon-nucleon cross sections}
In the literature, there are different fits to the free (total) nucleon-nucleon cross sections, such as those in Refs. 
\cite{BC10,JTWT93}. In this work, we have used the parametrization from the Ref. \cite{BC10} which is
obtained using the experimental data from Particle Data Group  \cite{pdgxnn}. 
For practical reasons, the free nucleon-nucleon cross sections are separated in three energy intervals, 
by means of the expressions
\begin{equation}
\sigma_{pp}=\left\{
\begin{array}
[c]{c}%
19.6+{4253/ E} -{ 375/ \sqrt{E}}+3.86\times 10^{-2}E \\
({\rm for }\ E < 280\  {\rm MeV}) \\ \; \\
32.7-5.52\times 10^{-2}E+3.53\times 10^{-7}E^3  \\
-  2.97\times 10^{-10}E^4  \\
({\rm for }\   280\ {\rm MeV}\le E < 840\  {\rm MeV}) \\ \; \\
50.9-3.8\times 10^{-3}E+2.78\times 10^{-7}E^2 \\
 +1.92\times 10^{-15} E^4  \\
({\rm for}\  840 \ {\rm MeV} \le E \le 5 \ {\rm GeV})
\end{array}
\right.
\label{signn1}
\end{equation}
for proton-proton collisions, and
\begin{equation}
\sigma_{np}=
\left\{
\begin{array}
[c]{c}%
89.4-{2025/ \sqrt{E}}+{19108/ E}-{43535/ E^2}
\\
 ({\rm for }\ E < 300\  {\rm MeV}) \\ \; \\
14.2+{5436/ E}+3.72\times 10^{-5}E^2-7.55\times 10^{-9}E^3
 \\
 ({\rm for }\   300\ {\rm MeV}\le E < 700\  {\rm MeV}) \\ \; \\
33.9+6.1\times 10^{-3}E-1.55\times 10^{-6}E^2 \\
 +1.3\times 10^{-10}E^3\\
 ({\rm for}\  700 \ {\rm MeV} \le E \le 5 \ {\rm GeV}) 
\end{array}
\right.
\label{signn2}
\end{equation}
for proton-neutron collisions. $E$ is the projectile laboratory
energy. The coefficients in the above equations have been obtained
by a least square fit to the nucleon-nucleon cross section
experimental data over a variety of energies, ranging from 10 MeV to
5 GeV. 

Most practical studies of medium corrections of nucleon-nucleon scattering are done by considering  
the effective two-nucleon interaction in infinite nuclear matter, or G-matrix,
as a solution of the Bethe-Goldstone equation \cite{GWW58}
\begin{equation}
\langle\mathbf{k}|\mathrm{G}(\mathbf{P},\rho_1,\rho_2)|\mathbf{k}_{0}\rangle
=\langle\mathbf{k}|\mathrm{v}_{NN}|\mathbf{k}_{0}\rangle
-\int{\frac
{d^{3}k^{\prime}}{(2\pi)^{3}}}{\frac{\langle\mathbf{k}|\mathrm{v}%
_{NN}|\mathbf{k^{\prime}}\rangle Q(\mathbf{k^{\prime}},\mathbf{P}%
,\rho_1,\rho_2)\langle\mathbf{k^{\prime}}|\mathrm{G}(\mathbf{P},\rho_1,\rho_2)|\mathbf{k}%
_{0}\rangle}{E(\mathbf{P},\mathbf{k^{\prime}})-E_{0}-i\epsilon}}\,
\label{10}%
\end{equation}
with $\mathbf{k}_{0}$, $\mathbf{k}$, and $\mathbf{k^{\prime}}$ the
initial, final, and intermediate relative momenta of the NN pair,
${\bf k}=({\bf k}_1-{\bf k}_2)/2$ and ${\bf P}=({\bf k}_1+{\bf k}_2)/2$. If energy and momentum is 
conserved in the binary collision, ${\bf P}$ is conserved in magnitude and direction, and the magnitude 
of ${\bf k}$ is also conserved. $\mathrm{v}_{NN}$ is the nucleon-nucleon potential. $E$ is the energy 
of the two-nucleon system, and $E_{0}$ is the same quantity on-shell. Thus
$
E(\mathbf{P},\mathbf{k})=e(\mathbf{P}+\mathbf{k})+e(\mathbf{P}-\mathbf{k}%
)$, with $e$ the single-particle energy in nuclear matter. It is also implicit in  Eq. \eqref{10} that 
the final momenta ${\bf k}$ of the NN-pair also lie outside the  range of occupied states. 

Eq.~(\ref{10}) is density-dependent due to the presence of the Pauli projection operator $Q$, defined by
\begin{equation}
Q(\mathbf{k},\mathbf{P},\rho_1,\rho_2)=\left\{
\begin{array}
[c]{c}%
1,\ \ \ \mathrm{if}\ \ \ k_{1,2}>k_{F1,F2}\\
0,\ \ \ \ \ \mathrm{otherwise.}%
\end{array}
\right. 
\label{12}%
\end{equation}
with $k_{1,2}$ the magnitude of the momenta of each nucleon. $Q$
prevents scattering into occupied intermediate states. The Fermi momenta $k_{F1,F2}$ are related to 
the proton and neutron densities by means of the zero temperature density approximation, $k_{Fi}=(3\pi^2 \rho_i/2)^{1/3}$. 
For finite nuclei, one usually replaces $\rho_i$ by the local densities to obtain the local Fermi 
momenta. This is obviously a rough approximation, but very practical and extensively used in the literature.
\begin{figure}[!t]
\center
\includegraphics[totalheight=8.0cm, angle=-90]{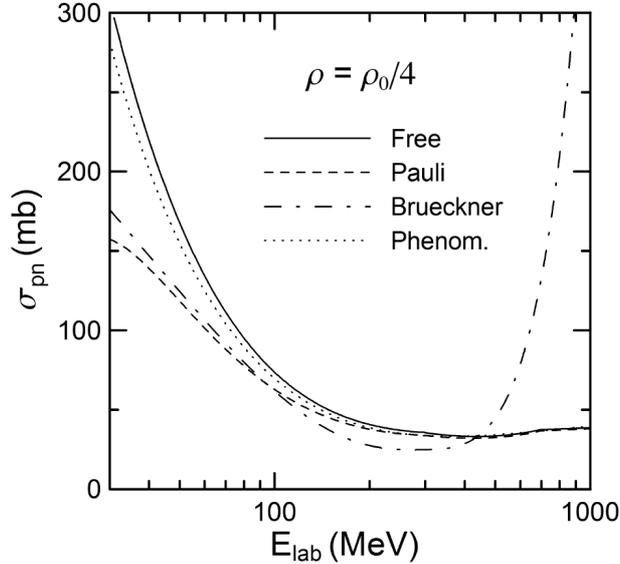}
\vspace{5mm} \caption{Parameterizations of proton-neutron cross sections as a function of the laboratory 
energy~\cite{BC10}. The solid line is the parametrization  of the free $\sigma_{pn}$ cross section given by Eq. \eqref{signn2}.  
The other curves include medium effects for symmetric nuclear matter for $\rho=\rho_0/4$, where $\rho_0=0.17$ fm$^{-3}$. 
The dashed curve includes the geometrical effects of Pauli blocking, as described by Eq.  \eqref{VM1}. 
The dashed-dotted curve is the result of the Brueckner theory, Eq. \eqref{brueckner}, and the dotted curve 
is the phenomenological parametrization, Eq. \eqref{pheno}.} 
\label{signpe}
\end{figure}
Only by means of several approximations, Eq. \eqref{10} can be related to nucleon-nucleon cross sections. 
If one neglects the medium modifications of the nucleon-mass, and scattering through intermediate 
states, the medium modification of the NN cross sections can be accounted for by the geometrical factor $Q$ only, that is,
\begin{equation}
\sigma_{NN}(k,\rho_1,\rho_2)=\int {\frac{d\sigma_{NN}^{free}}{d\Omega}}  Q(k,P, \rho_1,\rho_2) d\Omega , 
\label{snngeo}
\end{equation}
where $Q$ is now a simplified geometrical condition on the available scattering angles for the scattering 
of the NN-pair to unoccupied final states.

After this point, the geometrical treatment of Pauli corrections can be performed using the isotropic NN 
scattering approximation because the numerical calculations can be largely simplified if we assume that 
the free nucleon-nucleon cross section is isotropic. In this  case, a formula which fits the numerical 
integration of the geometrical model  reads \cite{BC10} 
\begin{eqnarray}
\sigma_{NN}(E,\rho_p,\rho_t) &=&\sigma_{NN}^{free}(E)\frac{1}{1+{1.892\left(\frac{2\rho_<}{\rho_0}\right)\left(\frac{|\rho_p-\rho_t|}{\tilde{\rho}\rho_0}\right)^{2.75}}}\nonumber \\
&\times& 
\left\{
\begin{array}
[c]{c}%
\displaystyle{1-\frac{37.02 \tilde{\rho}^{2/3}}{E}}, \ \ \   {\rm if} \ \ E>46.27 \tilde{\rho}^{2/3}\\ \, \\
\displaystyle{\frac{E}{231.38\tilde{\rho}^{2/3}}},\ \ \ \ \  {\rm if} \ \ E\le 46.27 \tilde{\rho}^{2/3}\end{array}
\right.
\label{VM1}
\end{eqnarray}
where $E$ is the laboratory energy in MeV, $\tilde{\rho}=(\rho_p+\rho_t)/\rho_0$, $\rho_<={\rm min} (\rho_p,\rho_t)$, $\rho_{i=p, t}$ 
is the local density of nucleus $i$, and $\rho_0=0.17$ fm$^{-3}$.
\begin{figure}[!t]
\centering
\includegraphics[totalheight=8.0cm, angle=-90]{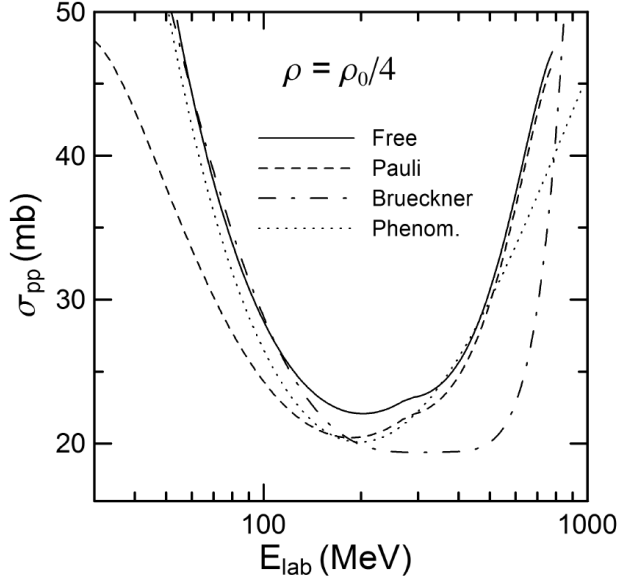}
\vspace{5mm} \caption{Same as in Figure \eqref{signpe}, but for pp collisions~\cite{BC10}.} 
\label{sigppe}
\end{figure}

The Brueckner method goes beyond this simple treatment of Pauli blocking, generating medium effects from  
nucleon-nucleon potentials,  such as the Bonn potential. An example is the work presented in Ref. \cite{LM:1993,LM:1994}, 
where a practical parametrization was given, which we will from now on refer as Brueckner theory. 
It reads\footnote{The misprinted factor 0.0256 in Ref. \cite{LM:1994} has been corrected to 0.00256.}
\begin{align}
\sigma_{np}  &  = \left[ 31.5 +0.092\left| 20.2-E^{0.53}\right|^{2.9}\right] \frac{1+0.0034E^{1.51} \rho^2}{1+21.55\rho^{1.34}} \nonumber\\
\sigma_{pp}  &  = \left[ 23.5 +0.00256\left( 18.2-E^{0.5}\right)^{4.0}\right] \frac{1+0.1667E^{1.05} \rho^3}{1+9.704\rho^{1.2}}.   \label{brueckner}
\end{align}
A modification of the above parametrization was done in Ref. \cite{Xian98}, which consisted in combining 
the free nucleon nucleon cross sections parametrized in Ref. \cite{Cha90} with the Brueckner theory 
results of Ref. \cite{LM:1993,LM:1994}. Their parametrization, which tends to reproduce better the 
nucleus-nucleus reaction cross sections, is 
\begin{align}
\sigma_{np} &= \left[ -70.67-18.18\beta^{-1}+25.26\beta^{-2}+113.85\beta\right]
\times \frac{1+20.88E^{0.04} \rho^{2.02}}{1+35.86\rho^{1.9}} \nonumber\\
\sigma_{pp} &= \left[ 13.73-15.04\beta^{-1}+8.76\beta^{-2}+68.67\beta^{4}\right]
\times \frac{1+7.772E^{0.06} \rho^{1.48}}{1+18.01\rho^{1.46}},   
\label{pheno}
\end{align}
where $\beta=\sqrt{1-1/\gamma^2}$ and $\gamma=E{\rm[MeV]}/931.5+1$.
We will denote Eq. \eqref{pheno} as the phenomenological parametrization.

The differences between the parametrization of the Brueckner, Eq.~\eqref{brueckner}, the geometrical 
Pauli blocking, Eq.~\eqref{VM1}, and the phenomenological one, Eq.~\eqref{pheno}, are visible. 
Figure \ref{signpe} is an example of that, where the varied parameterizations of proton-neutron 
cross sections are presented as a function of the laboratory energy. The solid line is the parametrization  
of the free $\sigma_{pn}$ cross section given by Eq. \eqref{signn2}. The other curves include medium effects 
for symmetric nuclear matter for $\rho=\rho_0/4$, where $\rho_0=0.17$ fm$^{-3}$. The dashed curve 
includes the geometrical effects of Pauli blocking, as described by Eq.  \eqref{VM1}. The dashed-dotted 
curve is the result of using the Brueckner theory, Eq. \eqref{brueckner}, and the dotted curve is the 
phenomenological parametrization, Eq. \eqref{pheno}. The large departure of results of the Brueckner 
parametrization above 300~MeV is not physical since Eq. \eqref{brueckner} is valid only under 300~MeV (pion production threshold) 
\cite{LM:1993,LM:1994}. On the other hand, the differences at lower energies are physical and 
Pauli-blocking effectively reduces the in-medium  {\it np} cross section. This is not so explicit 
in the phenomenological parametrization.

The above interpretation cannot be extended to the {\it pp} cross sections, which are shown in Figure 
\eqref{sigppe}. Here it is seen that the geometrical Pauli-blocking correction decreases the cross section  much more 
than in the other cases. Some important differences are also clearly visible at larger energies, $E\gtrsim 100$ MeV/nucleon. We now study the impact of these different methods on direct reactions at intermediate energies.

\subsection{Total reaction cross-sections}
As we mentioned before, obtaining a valid optical potential in knockout reactions~\cite{HRB91} and 
various direct reactions is crucial. One way to test the optical potentials is to reproduce total reaction cross
sections. As elastic scattering data at intermediate energies are scarce, for knockout reactions a proper test this can be done by calculating total reaction cross sections  for the core and the valence particle, separately. 
The total reaction cross-sections can be obtained in the framework of the eikonal approximation  as follows 
\begin{equation}
\sigma_R=2\pi \int db \ b \left[ 1- \left| S(b)\right|^2 \right],
\end{equation}
where $S$ is the eikonal $S$-matrices. The relation between optical potentials and
$S$-matrices is given by
\begin{equation}
S_i(b)=\exp[i\chi(b)]=\exp\left[-\frac{i}{\hbar v}\int_{-\infty}^\infty U_{iT}({\bf r})dz\right], \label{sib}
\end{equation} 
where $r=\sqrt{b^2+z^2}$, and $U_{iT}$ is the particle($i$)-target($T$) optical potential. A semiclassical 
probabilistic approach has been followed to calculate the cross sections and other observables in direct 
reactions as described in Refs. \cite{HM85,HBE96}, and a relation has been established between the 
optical potential and the nucleon-nucleon scattering amplitude in Ref. \cite{HRB91}. This relation 
is frequently  mentioned in the literature as the ``t-$\rho\rho$ approximation". 
``Experimentally deduced" optical potentials are often not available from elastic and inelastic scattering 
involving radioactive nuclei. Therefore, the t-$\rho\rho$ approximation is one of the most practical
techniques to obtain optical potentials. In this approximation, the eikonal phase becomes
\begin{equation}
\chi(b)=\frac{1}{k_{NN}}\int_{0}^{\infty}dq\ q\
\rho_{p}\left(  q\right) \rho_{t}\left(  q\right)  f_{NN}\left(
q\right)  J_{0}\left(  qb\right)
\ ,\label{eikphase}%
\end{equation}
where $\rho_{p,t}\left(  q\right)  $ is the Fourier transform of the nuclear
densities of the projectile and target, and $f_{NN}\left(  q\right)  $ is the
high-energy nucleon-nucleon scattering amplitude at forward angles, which can
be parametrized as
\begin{equation}
f_{NN}\left(  q\right)  =\frac{k_{NN}}{4\pi}\sigma_{NN}\left(  i+\alpha
_{NN}\right)  \exp\left(  -\beta_{NN}q^{2}\right)  \ .
\label{fnn}%
\end{equation}
\begin{figure}[b]
\center
\includegraphics[scale=0.4,keepaspectratio=true,clip=true]{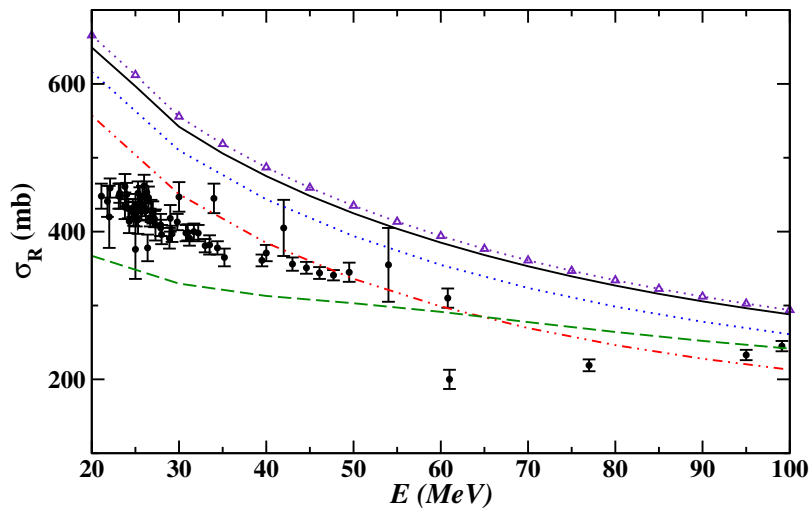}
\caption[totsigR]{The total reaction cross section of the p\;+$\;^{12}$C taken from 
Ref.~\cite{CAR96}. The curves are calculated with the free NN cross sections from Ref.~\cite{JTWT93} 
(solid), with a geometrical account of Pauli blocking (dashed), a phenomenological fit from 
Ref.~\cite{Xian98} (dotted), and a correction from Brueckner theory (dashed-dotted).
The triangle-dotted curve is calculated with the same free NN cross sections as in Ref.~\cite{JTWT93}, 
but with an another HFB calculation \cite{Bei74} for the $^{12}$C ground state density.}
\label{totsigR}
\end{figure}
One often neglects nuclear medium effects in the experimental analysis of knockout reactions, as pointed out in Ref. \cite{BC10}.
However, their importance has been well known for a long time in the study of elastic and inelastic 
scattering, as well as of total reaction cross sections \cite{HRB91,Ra79}. 
In these situations, a systematic analysis of the medium effects has been presented in Ref. \cite{HRB91}, and it has shown
that the effects becomes larger at lower energies, where Pauli blocking strongly reduces the 
nucleon-nucleon cross sections in the medium. 

The p + $^{12}$C total reaction cross sections in the energy range of 20-100 MeV/nucleon shown in Figure \ref{totsigR} presents the justification 
of these statements,  where the 
experimental data taken from the Ref.~\cite{CAR96}. The cross sections were calculated from the Eqs. (\ref{sib},\ref{eikphase},\ref{fnn}) 
and $^{12}$C density from a Hartree-Fock-Bogoliubov calculation (HFB) \cite{Bei74}.  
Various different calculations are shown in Figure \ref{totsigR}.
The result of Eq. \eqref{eikphase} with the free nucleon-nucleon cross sections and the carbon matter 
density from a HFB calculation \cite{Nus12} is represented by the solid curve, whereas the triangle-dotted  
curve (the triangles are not data, but used for better visibility) uses a different HFB density \cite{Bei74},  
consistent with the calculations presented in Ref. \cite{trache02}. As expected, that the agreement between 
the two calculations is very good. 

The same calculation procedure, but this time including medium corrections for the nucleon-nucleon 
cross section, has been performed to obtain the other curves in Figure \ref{totsigR}.
It is evident that the results are very different than the former.
The medium effects with various different are shown with the dotted, dashed-dotted, 
and dashed curves, which they correspond to the calculations with phenomenological, Brueckner, and 
Pauli geometrical methods, respectively. Obviously, medium effects modify the results, yielding a 
closer reproduction of the data. But the large experimental error bars do not allow a fair judgment of 
which model reproduces better the data.

\subsection{Nucleon knockout reactions}
Momentum distributions of the projectile-like residues in one-nucleon knockout are a measure of 
the spatial extent of the wavefunction of the struck nucleon, while the cross section for the nucleon 
removal scales with the occupation amplitude, or probability (spectroscopic factor), for the given 
single-particle configuration in the projectile ground state. 
The longitudinal momentum distributions are given by (see, e.g., Refs. \cite{HBE96,BH04,BG06}) 
\begin{align}
\frac{d\sigma_{\mathrm{str}}}{dk_{z}}  &  = (C^2S) \frac{1}{2\pi%
}\frac{1}{2l+1}\sum_{m}\int_{0}^{\infty}d^{2}b_{n} \left[  1-\left\vert
S_{n}\left(  b_{n}\right)  \right\vert ^{2}\right]  \nonumber\\
&  \times  \int_{0}^{\infty}%
d^{2}b_c \ \left\vert S_{c}\left(  b_{c}\right)  \right\vert ^{2}\left\vert \int_{-\infty}^{\infty}dz \exp\left[  -ik_{z}z\right]
\psi_{lm}\left(  \mathbf{r}\right)  \right\vert ^{2},\label{strL}%
\end{align}
where $k_{z}$ represents the longitudinal component of $\mathbf{k}_{c}$ (final momentum of the core 
of the projectile nucleus) and $(C^2S)$ is the spectroscopic factor, and $\psi_{lm}\left(\mathbf{r}\right)$
is the wavefunction of the core plus (valence) nucleon system $(c+n)$ in a state with 
single-particle angular momentum $l,m$.

\subsubsection{$^{12}${\rm C}($^{17}${\rm C},$^{16}${\rm B)X} at {\rm 35~MeV/u}\\}
\begin{figure}[b]
\center
\includegraphics[scale=0.4,keepaspectratio=true,clip=true]{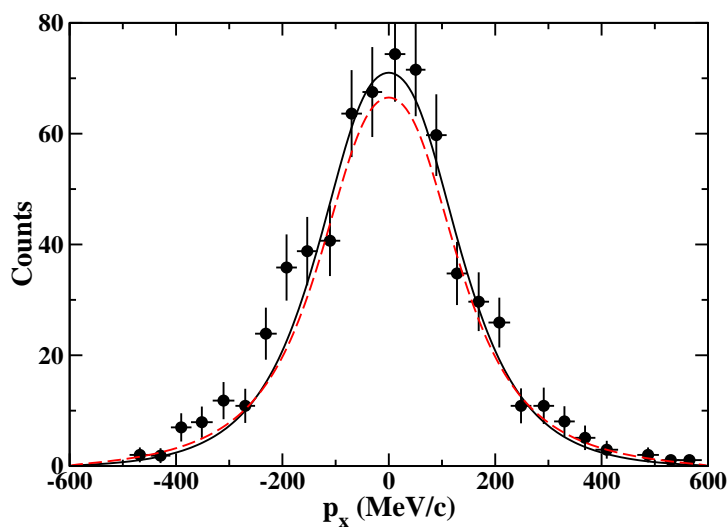}
\caption[c17c12]{Transverse momentum distributions for the $^{12}$C($^{17}$C,$^{16}$B)X system at 35~MeV/u. 
Solid lines represent calculations including medium corrections. Dashed lines stem from calculations 
that do not include medium corrections. The data are taken from Ref. \cite{LEC09}.}
\label{c17c12}
\end{figure}
The one-proton removal reaction, $^{12}$C($^{17}$C,$^{16}$B)X, from $^{17}$C at 35~MeV/nucleon has been 
measured with the aim to explore the low-lying structure of the unbound $^{16}$B nucleus. 
In Ref. \cite{LEC09}, the unbound $^{16}$B nucleus is  assumed to be  a $d$-wave neutron decay from 
$^{15}$B+$n$ system. Here, we have focused to study the consequences 
of medium corrections on the transverse momentum distribution of the $^{16}$B fragment following the 
same assumptions as in Ref. \cite{LEC09}. The configuration mixing of the proton removed from $^{17}$C 
is assumed to be
\begin{eqnarray} 
&&|^{17}\text{C}\rangle = \alpha_1|^{16}\text{B}(0^-)\otimes \pi 1p_{3/2}\rangle \nonumber \\
&+& \alpha_2|^{16}\text{B}(3^-_1)\otimes \pi1p_{3/2}\rangle + \alpha_3|^{16}\text{B}(2^-_1)\otimes \pi1p_{3/2}\rangle \nonumber \\ 
&+& \alpha_4|^{16}\text{B}(2^-_2)\otimes \pi1p_{3/2}\rangle + \alpha_5|^{16}\text{B}(1^-_1)\otimes \pi1p_{3/2}\rangle \nonumber \\ 
&+& \alpha_6|^{16}\text{B}(3^-_2)\otimes \pi1p_{3/2}\rangle,
\end{eqnarray}
where $\alpha_i$ is the spectroscopic amplitude for a core-single particle configuration $i=(c\otimes nlj)$. 

In Ref.~\cite{LEC09} a good agreement between data and calculated transverse momentum distributions
was achieved using the spectroscopic factors from a shell-model calculation with the WBP interaction 
\cite{WAR92}. However, they have obtained a theoretical result of 24.7~mb for total cross section 
which diverges from the measured cross-section, 6.5(1.5)~mb. In Ref. \cite{GAD04}, an explanation is 
proposed to this large discrepancy as due to a reduction of the spectroscopic factor by 70\%  for strongly bound 
nucleon systems. The theoretical estimates for the cross section with the reduction at the spectroscopic 
factor becomes 7.5~mb, in reasonable accordance with the data.
We do not challenge the assumptions of Ref. \cite{LEC09},
and we use the same configuration mixing and spectroscopic factors as in \cite{LEC09}. The proton binding 
potential parameters are given in Ref.~\cite{KBBT12}, which are adjusted to obtain the effective separation 
energies. Here, as it is shown in Figure~\ref{c17c12}, we find that medium corrections change the 
cross sections by 5\% which is rather small to explain the observed difference with the total cross sections. 

\subsubsection{$^9${\rm Be(}$^{11}${\rm Be,}$^{10}${\rm Be)X} at {\rm 60~MeV/u}\\}
\begin{figure}[b]
\center
\includegraphics[scale=0.4,keepaspectratio=true,clip=true]{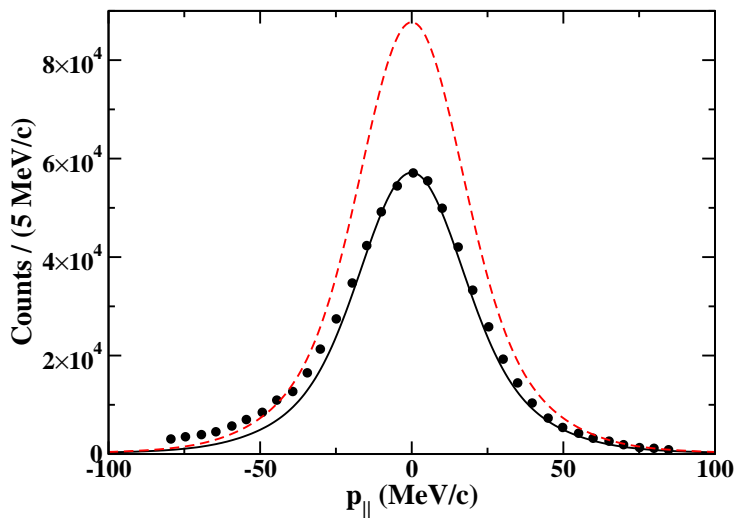}
\caption[be11be9]{Longitudinal momentum distributions of for the reaction $^9$Be($^{11}$Be,$^{10}$Be) 
at 60~MeV/nucleon. Solid lines represent calculations including medium corrections. Dashed lines stem 
from calculations that do not include medium corrections. The data is taken from Ref.~\cite{AUM00}.}
\label{be11be9}
\end{figure}
In order to further understand the medium effects on knockout reactions, we consider the $^9$Be($^{11}$Be,$^{10}$Be)X 
system at 60~MeV/u which can be modeled by a core plus valence system with the assumption $^{10}$Be$_{gs}(0^+)+{\rm n}$ 
in $2s_{1/2}$ orbital for the ground state of $^{11}$Be$_{gs}(1/2^+)$ ($S_n=0.504$~MeV). Here we use 
the same Woods-Saxon potential parameters for the bound state as published in Ref. \cite{HBE96}: ($R_0=2.70$ fm, $a_0=0.52$ fm). 
In Figure~\ref{be11be9} and Table \ref{cross} we present our results for the the neutron removal longitudinal 
momentum distribution of 60~MeV/nucleon $^{11}$Be projectiles incident on $^9$Be targets. We find that 
medium corrections for this system change the cross sections by 50\% which is quite big. 

It has been show  that $^{17}$C has a small ``effective" size and that $^{11}$Be
has a large effective size among the low energy systems studied in Ref. \cite{KBBT12}. Therefore, the wavefunctions of weakly  
bound systems extend far within the target where the nucleon-nucleon cross sections are strongly modified 
by the medium. Momentum distributions and nucleon removal cross sections in knockout reactions are thus 
expected to change appreciably with the inclusion of medium corrections of nucleon-nucleon cross section. 
Such corrections are also expected to play a more significant role for loosely-bound systems.

\begin{table}[hbtp]
\begin{center}
\setlength{\tabcolsep}{.25cm}
\begin{tabular}{|c|cc|cc|} 
\hline
$\sigma_{-1n}$&\multicolumn{2}{c|}{$^{12}$C($^{17}$C,$^{16}$B)}&\multicolumn{2}{c|}{$^9$Be($^{11}$Be,$^{10}$Be)}\\
$=\sigma_{dif}+\sigma_{str}$ &  Full & no medium  &   Full& no medium \\
\hline
Strip. [mb]                  &  7.56 &    5.63 & 122.5 &  164.1  \\ 
Diff.  [mb]                  & 18.42 &   19.15 &  49.6 &   97.3  \\ 
Total  [mb]                  & 25.98 &   24.78 & 172.1 &  261.4  \\
\hline
\end{tabular}
\caption{The cross sections calculated for the systems, $^{12}$C($^{17}$C,$^{16}$B) at 35~MeV/nucleon and 
$^9$Be($^{11}$Be,$^{10}$Be) at 60~MeV/nucleon.}
\label{cross}
\end{center}
\end{table} 

\section{Summary}
In this small report, we have explored the importance of the medium modifications of the nucleon-nucleon 
cross sections on direct reactions, and particularly on knockout reactions. It has been shown
that the effects are noticeable at low energies. Nonetheless, we have noticed that medium effects do not lead to sizable 
modifications on the shapes of momentum distributions. 
We have shown this explicitly by comparing our results with a large number of available 
experimental data in Ref.~\cite{KBBT12}. As expected on physics grounds, these corrections are  larger 
for experiments at lower energies, around 50~MeV/nucleon, and for weakly bound nuclei. 

Medium effects in knockout reactions have been frequently ignored in the past. We show that they 
have to be included in order to obtain a better accuracy of the extracted spectroscopic factors. 
Although these conclusions might not come as a big surprise, they have not been properly included in 
many previous experimental analyses.  

\ack
This work was partially supported by the US-DOE grants DE-SC004972 and DE-FG02-08ER41533 and 
DE-FG02-10ER41706, and by the Research Corporation.

\end{document}